\documentclass[english,aps,prl,superscriptaddress,showpacs,twocolumn]{revtex4}
\usepackage[T1]{fontenc}
\usepackage[latin9]{inputenc}
\usepackage{graphicx}

\makeatletter

\usepackage{bm}

\makeatletter
\makeatother

\makeatother

\usepackage{babel}

\begin{document}

\title{Effects of nematic fluctuations on the elastic properties of iron
arsenide superconductors}

\author{R. M. Fernandes$^{1}$, L. H. VanBebber$^{2}$, S. Bhattacharya$^{3}$,
P. Chandra$^{4}$, V. Keppens$^{2}$, D. Mandrus$^{5}$,
M. A. McGuire$^{5}$, B. C. Sales$^{5}$, A. S. Sefat$^{5}$
\& J. Schmalian }

\affiliation{Ames Laboratory \& Department of Physics and Astronomy, Iowa State
Univ., Ames, IA 50011, USA \\
 $^{2}$Department of Materials Science, University of Tennessee,
Knoxville, TN 37996, USA \\
 $^{3}$Tata Institute of Fundamental Research 400005, Mumbai, India
\& Cavendish Laboratory, Univ. of Cambridge CB3 0HE, UK\\
 $^{4}$Department of Physics and Astronomy \& Center for Emergent
Materials, Rutgers Univ., Piscataway, NJ 08855, USA \\
 $^{5}$Materials Science and Technology Division, Oak Ridge National
Laboratory, Oak Ridge, TN 37831, USA}

\date{\today}

\begin{abstract}
We demonstrate that the changes in the elastic properties of the FeAs
systems, as seen in our resonant ultrasound spectroscopy data, can
be naturally understood in terms of fluctuations of emerging nematic
degrees of freedom. Both the softening of the lattice in the normal,
tetragonal phase as well as its hardening in the superconducting phase
are consistently described by our model. Our results confirm the view
that structural order is induced by magnetic fluctuations.
\end{abstract}


\maketitle
In the recently discovered FeAs superconductors \cite{Kamihara08,Rotter08},
the rather large superconducting transition temperatures are in close
proximity to an antiferromagnetic (AFM) phase transition at $T_{\mathrm{N}}$
and a structural transition from tetragonal (Tet) to orthorhombic
(Ort) at $T_{\mathrm{S}}$ \cite{Luetkens,Ni09,Chu09,Ning09}. This
begs the question to what extent spin and lattice degrees of freedom
play important roles for high-$T_{c}$ superconductivity (SC). The
fact that $T_{\mathrm{S}}$ and $T_{\mathrm{N}}$ track each other
closely (see inset of Fig.3), as well as the existence of an isotope
effect for $T_{\mathrm{N}}$ \cite{Liu09}, are clear evidence 
that they signal the onset of strongly
coupled ordered states. Since $T_{\mathrm{S}}\geq T_{\mathrm{N}}$, it seems, at
first glance, plausible to assume that lattice effects are primary
and the AFM order is simply a secondary effect. However, it has
been suggested \cite{Fang08,Xu08} that spin fluctuations, related
to the observed AFM ordering at $T_{\mathrm{N}}$, will, at higher
temperatures, lead to emergent nematic degrees of freedom \cite{chandra}
that couple to the lattice \cite{Qi09,Gorkov09}. Even though of
magnetic origin, probing nematic fluctuations is nontrivial, since
the magnetic field fluctuations associated to them average to zero.
Thus, determining to what extent these fluctuations are present 
is key for identifying the relevant low energy modes
in the pnictides.

\begin{figure}

\begin{centering}
\includegraphics[width=0.9\columnwidth]{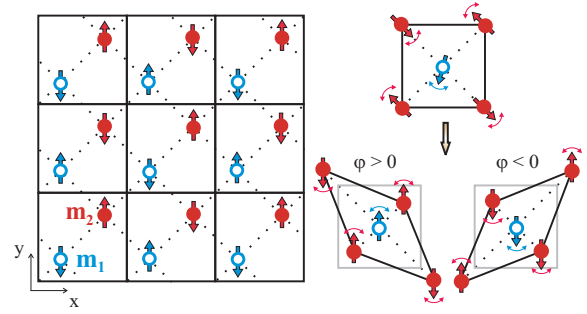} 
\par\end{centering}

\caption{\textbf{(left panel)} In-plane magnetic ordering of the pnictides.
The two iron atoms present in one tetragonal unit cell (open blue
circles and filled red circles) form two coupled AFM sublattices with
co-planar staggered magnetization $\mathbf{m}_{1}$ and $\mathbf{m}_{2}$.
\textbf{(right panel)} In the disordered phase, the spins in the two
sublattices fluctuate around two arbitrary staggered magnetization
directions. In the nematic phase, these two N\'eel vectors are locked
either parallel ($\varphi>0$) or anti-parallel ($\varphi<0$) to
each other, but still with $\left\langle \mathbf{m}_{1}\right\rangle =\left\langle \mathbf{m}_{2}\right\rangle =0$.
The coupling between the nematic and the elastic order parameters
makes the bonds between neighboring parallel (antiparallel) spins
contract (expand).}

\end{figure}

Recent experiments \cite{Chuang10, Chu10} found strong anisotropy in the electronic properties of the Ort phase,
suggesting an underlying electronic nematic state.
In this Letter, we focus on the Tet phase and demonstrate that the measurement of the shear modulus
$C_{\mathrm{s}}$ allows for a direct determination of the nematic
order-parameter susceptibility $\chi_{\mathrm{nem}}$. 
In particular, we show that the high-temperature shear
modulus $C_{s,0}$ is renormalized by nematic fluctuations to:

\begin{equation}
C_{\mathrm{s}}^{-1}=C_{\mathrm{s},0}^{-1}+\lambda^{2}C_{\mathrm{s},0}^{-2}\chi_{\mathrm{nem}},\label{shearmodulus}\end{equation}
where $\lambda$ is the magneto-elastic coupling. We calculate $\chi_{\mathrm{nem}}$
in the Tet phase using a $1/N$ approach and compare the results with
Resonant Ultrasound Spectroscopy (RUS), which measures $C_{\mathrm{s}}$.
Our RUS data show that nematic fluctuations are prominent over a large
portion of the phase diagram of the pnictides. Although driven
by AFM fluctuations, they represent new emergent collective degrees
of freedom. 
Their coupling to the lattice makes the structural transition coincident with the nematic transition,
implying that the elastic degrees of freedom are secondary with respect to the electronic ones.
Here, we analyze the behavior of nematic fluctuations not only in
the normal state, but also in the vicinity of $T_{c}$, explaining
the observed hardening of $C_{\mathrm{s}}$ in the SC state.

In the AFM phase, spins are coupled antiferromagnetically for Fe-Fe
bonds along one diagonal and ferromagnetically along the other diagonal
of the two-Fe unit cell (Fig. 1). The in-plane magnetic ordering can be divided
into two sublattices \cite{Yildirim08} with N\'eel magnetizations
$\mathbf{m}_{1}$ and $\mathbf{m}_{2}$, which are weakly coupled
since in the classical ground state, local
fields associated with one sublattice cancel in the other. 

This sublattices coupling is related to a $Z_{2}$ (i.e. discrete,
Ising type) symmetry: $\mathbf{m}_{1}\rightarrow-\mathbf{m}_{1}$,
$\mathbf{m}_{2}\rightarrow\mathbf{m}_{2}$ along with a rotation of
the coordinates by $\pi/2$. The $Z_{2}$ symmetry, manifest by the
two possible orientations of the magnetic stripes, will give rise
to an Ising-like order parameter, which in the continuous limit is given by
$\varphi\left(x\right)=\mathbf{m}_{1}\left(x\right)\cdot\mathbf{m}_{2}\left(x\right)$
\cite{chandra}, with $x=\left(\mathbf{r,}\tau\right)$ as shown
in Fig. 1. It describes the relative orientation of $\mathbf{m}_{1}$ and $\mathbf{m}_{2}$
and does not change sign upon magnetic field inversion. The particular
magnetic structure of the pnictides naturally accounts for this
emergent degree of freedom, regardless of whether one considers an itinerant
\cite{Eremin09,schmalianmazin} or a localized \cite{Si08} picture.
Once $\left\langle \varphi\right\rangle \neq0$, the
$Z_{2}$ symmetry is spontaneously broken \cite{chandra},
and so is the underlying $C_4$ symmetry of the lattice.
This is why one associates the Ising-ordered phase to a "nematic" state, 
even though, strictly speaking, the full rotational symmetry, as well as the translational symmetry, 
are already broken above $T_{\mathrm{S}}$.

To probe $\varphi$, we use the fact that the nematic order couples to the elastic
degrees of freedom of the lattice \cite{Fang06,Qi09}:
in orthorhombic fluctuations, present for $T>T_{\mathrm{S}}$, bonds
connecting antiparallel spins are expanded whereas those connecting
parallel spins contract \cite{zhao,geibel} (see Fig. 1). There
is therefore an energy cost associated with such an orthorhombic fluctuation,
characterized by the shear modulus $C_{\mathrm{s}}\equiv C_{66}$.
Shear strain ($\epsilon_{\mathrm{s}}=2\varepsilon_{xy}$, where $\epsilon_{\alpha\beta}$
is the strain tensor) and the nematic order parameter are linearly
coupled in the energy: \begin{equation}
S_{\mathrm{el-mag}}=\lambda\int_{x}\epsilon_{s}\left(x\right)\varphi\left(x\right).\label{shearcoupling}\end{equation}
 i.e. they order simultaneously. Here $\lambda>0$ is the magneto-elastic
coupling constant and $\int_{x}...=\int_{0}^{T^{-1}}d\tau\int d^{3}r...$.
Assuming a harmonic lattice for $\lambda=0$, with bare (high-temperature) shear modulus $C_{\mathrm{s},0}$,
we obtain the renormalized $C_{\mathrm{s}}$ in
Eq. (\ref{shearmodulus}). Note that the static nematic susceptibility
$\chi_{\mathrm{nem}}=\chi_{\mathrm{nem}}\left(q\rightarrow0\right)$
is a four-spin correlation function, given by $\chi_{\mathrm{nem}}\left(q\right)=\left\langle \varphi_{q}\varphi_{-q}\right\rangle -\left\langle \varphi_{q}\right\rangle ^{2}$,
where $\varphi_{q}$ denotes the Fourier transform of $\varphi\left(x\right)$.
From Eq.\ref{shearmodulus}, we see that nematic fluctuations soften
the lattice in the Tet phase ($T>T_{\mathrm{S}}$). A divergence
of $\chi_{\mathrm{nem}}$ brings $C_{\mathrm{s}}$ to zero, signaling
the onset of the Tet-Ort transition.

To probe the shear modulus in the pnictides, RUS \cite{migliori}
was performed on undoped BaFe$_{2}$As$_{2}$ and optimally doped
BaFe$_{1.84}$Co$_{0.16}$As$_{2}$ single crystals grown out of FeAs
flux \cite{sefat}. More than $20$ mechanical resonance frequencies
$f$ were detected. Since the crystals have $6$ independent elastic
constants $C_{ij}$ in the Tet phase, it is clear that each resonance
line $f$ will be proportional to a combination of $C_{ij}$. Extraction
of the full elastic tensor is not possible due to samples size limitations,
but information about $C_{s}=C_{66}$ can be obtained. Because the
samples undergo a Tet-Ort transition, $C_{s}$ must get soft by symmetry.
Numerical simulations of the resonance spectrum show that this soft
$C_{s}$ dominates several of the measured lines, with $f^{2}\propto C_{s}$
\cite{migliori}. 

Following this procedure we present (Fig. 2) the $T$ dependence of
two resonance frequencies that are representative of a larger set
of $C_{s}$-dominated lines. The observed reduction in $f^{2}$ indicates
a dramatic softening of $C_{s}$. In the undoped case (Fig. 2a), this
softening is cut off by a simultaneous structural/AFM weak first order
transition at $T_{\mathrm{S}}=130\mathrm{K}$. In the optimally doped
case (Fig. 2b), where no AFM or structural transitions are present,
the softening continues to lower temperatures until it is truncated
by the SC transition at $T_{\mathrm{c}}=22\mathrm{K}$. In order to
confirm that these features are not due to any particular mechanical
instability of the samples, we also performed RUS in strongly doped
samples that present neither SC nor structural transitions. In this
case, only an uniform, mild increase in $f$ with decreasing $T$
was observed.

\begin{figure}
\begin{centering}
\includegraphics[width=0.8\columnwidth]{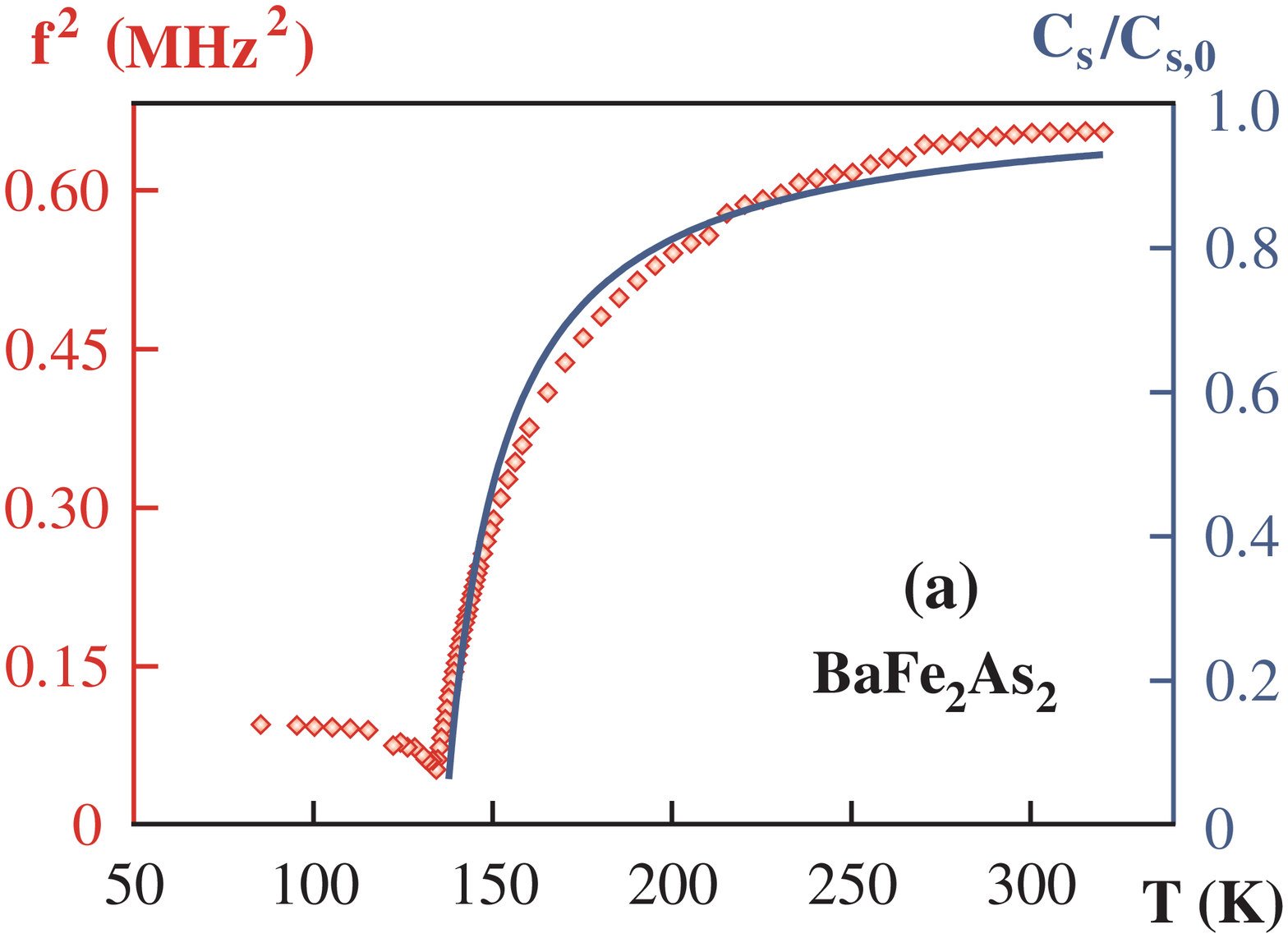} 
\par\end{centering}

\begin{centering}
\bigskip{}

\par\end{centering}

\begin{centering}
\includegraphics[width=0.8\columnwidth]{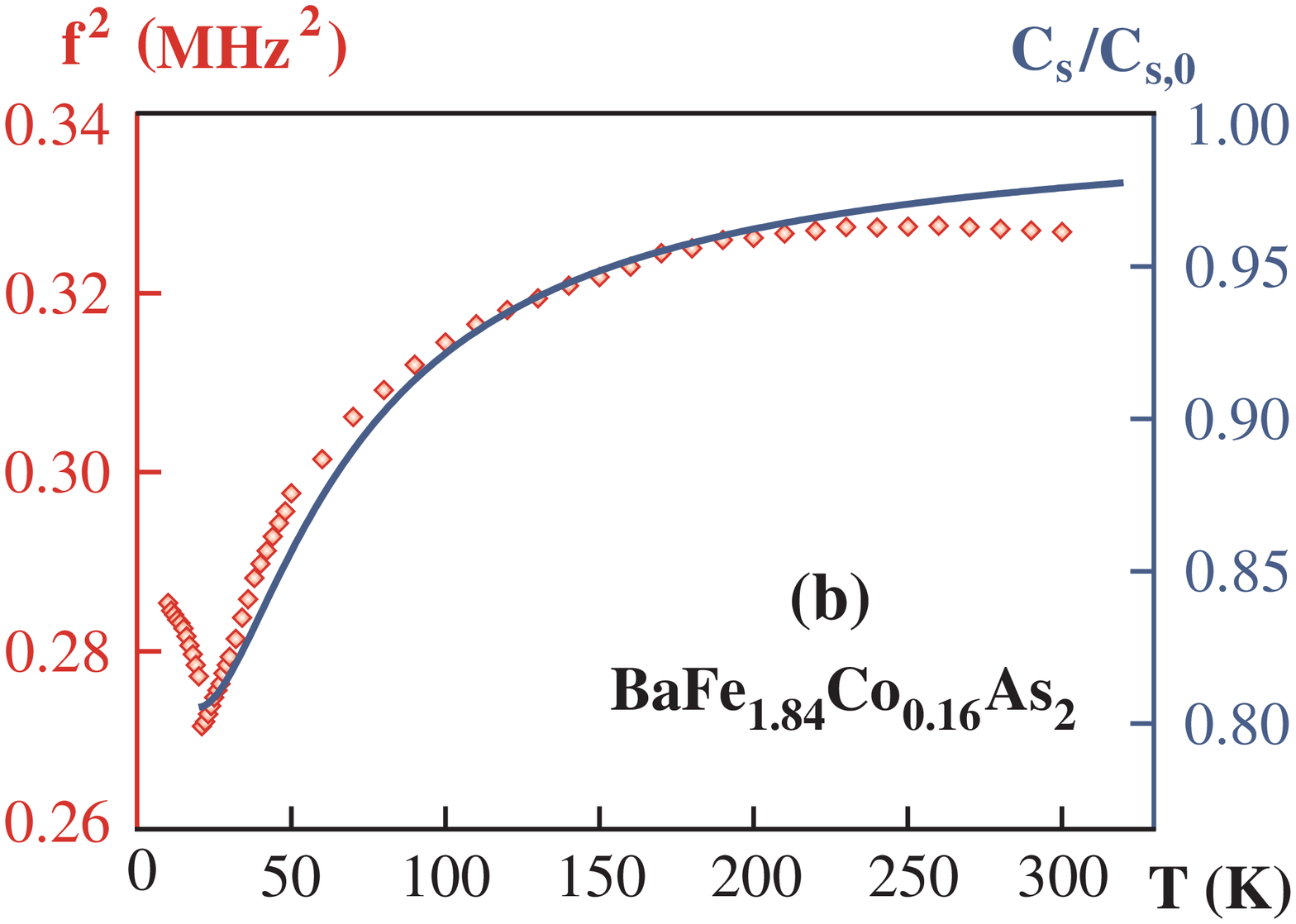} 
\par\end{centering}

\caption{Temperature dependence of the RUS measured squared resonant frequency
$f^{2}$ (red points), and of the calculated shear modulus $C_{s}\equiv C_{66}$
of the tetragonal phase (solid blue line), for both \textbf{(a)} undoped
BaFe$_{2}$As$_{2}$ and \textbf{(b)} optimally doped BaFe$_{1.84}$Co$_{0.16}$As$_{2}$
.}

\end{figure}

Next, we calculate the nematic susceptibility $\chi_{\mathrm{nem}}$
in Eq.\ref{shearmodulus}. Following Refs. \cite{Fang08,Xu08} we
introduce an action $S=S_{\mathrm{mag}}+S_{\mathrm{el}}+S_{\mathrm{el-mag}}$
for the magnetic and elastic degrees of freedom, with $S_{\mathrm{el-mag}}$
given by Eq.\ref{shearcoupling}. For each sublattice in Fig. 1, we
consider: \begin{equation}
S_{\mathrm{mag}}^{\left(0\right)}\left[\mathbf{m}_{i}\right]=\frac{1}{2}\int_{q}\tilde{\chi}_{0,q}^{-1}\left\vert \mathbf{m}_{i}\left(q\right)\right\vert ^{2}+\frac{w}{4}\int_{x}\mathbf{m}_{i}^{4}\left(x\right),\label{S0_mag}\end{equation}
where $w>0$ is a coupling constant and $q=\left(\mathbf{q,}i\omega_{n}\right)$
refers to momentum, $\mathbf{q}$, measured relative to the ordering
vector $\mathbf{Q}=\left(\frac{\pi}{a},\pm\frac{\pi}{a}\right)$,
and bosonic Matsubara frequency $\omega_{n}$, with $\int_{q}...=T\sum_{n}\int\frac{d^{3}q}{\left(2\pi\right)^{3}}...$.
The bare spin susceptibility $\tilde{\chi}_{0,q}$ is assumed to be
$\tilde{\chi}_{0,q}^{-1}=r_{0}+\mathbf{q}_{\parallel}^{2}-2J_{z}\cos\left(q_{z}c\right)+\Pi\left(\omega_{n}\right)$.
Here, the particle-hole bubble $\Pi\left(\omega_{n}\right)$ is determined
by the electronic structure and leads to a sensitivity of $\tilde{\chi}_{0,q}$
with respect to SC. Not only the static part of $\tilde{\chi}_{0,q}$
is changed in the SC state, but also its dynamics\cite{abanov}:
the opening of the SC gap $\Delta$ makes the spin dynamics ballistic
for $\omega_{n}\ll\Delta$ ($z=1$), while for $\omega_{n}\gg\Delta$
one recovers the overdamped dynamics of a metallic AFM ($z=2$). Thus,
we can write $\Pi\left(\omega_{n}\right)-\Pi\left(0\right)=\Gamma_{z}\left\vert \omega_{n}\right\vert ^{2/z}$
with Landau particle-hole damping coefficient $\Gamma_{z=2}=\gamma^{-1}$
as well as $\Gamma_{z=1}\simeq\left(\Delta\gamma\right)^{-1}$. $\ r_{0}$
characterizes the distance to a magnetic critical point and $J_{z}$
is a magnetic coupling between Fe-As layers separated by $c$. To
describe systems of different Co doping, we vary $\gamma$ and $r_{0}$.
The full magnetic action is given by the sum of the contributions
of the two sublattices and the dominant \cite{Xu08,Qi09} term that
couples them: \begin{equation}
S_{\mathrm{mag}}=\sum_{i}S_{\mathrm{mag}}^{\left(0\right)}\left[\mathbf{m}_{i}\right]-\frac{g}{2}\int_{x}\left(\mathbf{m}_{1}\left(x\right)\cdot\mathbf{m}_{2}\left(x\right)\right)^{2}\label{S_mag}\end{equation}
where $g>0$ is the sublattice coupling. In a localized picture, $g$
is related to quantum and thermal spin fluctuations \cite{chandra},
whereas in an itinerant picture it is related to the ellipticity of
the electron band \cite{Eremin09,Antropov08}. To solve for the coupled
magnetic and elastic system we assume that $\mathbf{m}_{i}$ is an
$N$-component vector and analyze the leading term for large-$N$
\cite{Fang08,Xu08}. We find: \begin{equation}
\chi_{\mathrm{nem}}\left(q\right)=\frac{\chi_{0,\mathrm{nem}}\left(q\right)}{1-g_{r}\chi_{0,\mathrm{nem}}\left(q\right)}\end{equation}
where $g_{r}=g+\lambda^{2}/C_{s,0}$ is the sublattice coupling renormalized
by magneto-elastic interactions. Thus, even if $g=0$, the magneto-elastic
coupling gives an effective coupling between the sublattices. $\chi_{0,\mathrm{nem}}\left(q\right)=N\int_{p}\tilde{\chi}_{p}\tilde{\chi}_{p+q}$
is determined by the dynamic magnetic susceptibility $\tilde{\chi}_{p}$
of Eq.\ref{S0_mag} but with $r_{0}+\Pi\left(0\right)$ replaced by
$r=\xi^{-2}$. $\xi$ is the magnetic correlation length, determined
in the large $N$ approach. If $d+z\leq4$ it follows that $\chi_{0,\mathrm{nem}}\left(0\right)$
diverges as $\xi\rightarrow\infty$. Here, the system has $d=2$ ($d=3$)
behavior for $r\gg J_{z}$ ($r\ll J_{z}$). Thus, a sufficiently
large, but finite $\xi$ suffices to cause a divergence of $\chi_{\mathrm{nem}}$,
implying that the Tet-Ort transition is induced by AFM fluctuations,
not AFM order; hence $T_{\mathrm{S}}>T_{\mathrm{N}}$ (except when
both transitions are first order).

First, we present our results for the Tet and non-SC state in Fig.
2, together with the RUS data. The spin dynamics is overdamped (i.e.
$z=2$), as seen in inelastic neutron scattering \cite{inosov}.
We used the parameters $\gamma_{\mathrm{undoped}}^{-1}=2.2\times10^{-3}$,
$\gamma_{\mathrm{doped}}^{-1}=4.7\times10^{-3}$, $g=0.015\epsilon_{0}$,
$\frac{\lambda^{2}}{C_{s,0}}=10^{-3}\epsilon_{0}$, and $J_{z}=10^{-3}\epsilon_{0}$,
where $\epsilon_{0}\approx3.4$ meV is a magnetic energy scale. The
value for the Landau damping is consistent with that used in Ref.
\cite{inosov}. The lattice softening, seen in the experiment over
a large temperature range, can be well described by the above theory,
suggesting nematic fluctuations up to room temperature and over a
significant portion of the phase diagram. The amplitude of the effect
indicates a strong magneto-elastic coupling,
consistent with the observed softening of some phonon frequencies
\cite{mittal}.

For both samples, the lattice softening in the RUS data stops at low $T$. 
For zero doping this happens at the joint first order
structural/AFM transition, where a hardening of the lattice is expected.
Due to the Fermi surface reconstruction in the AFM phase, however,
a quantitative description of $C_{s}$ below $T_{N}$ is beyond the
scope of our model.

For optimal doping, the lattice also gets harder below $T_{c}$. In
this case, though, there is a priori no reason for $C_{s}$ to be
minimum at $T_{\mathrm{c}}$. Note that this behavior is distinct
from the Testardi-behavior seen in other superconductors, where strain
couples to the squared SC order parameter, $\left\vert \Psi\right\vert ^{2}$,
and changes the transition temperature $T_{c}\rightarrow T_{c}\left(\varepsilon_{ab}\right)$,
making the elastic modulus behave very similar to minus the heat capacity
at the transition \cite{Millis88,Nyhus02}. Here, symmetry prohibits
such a coupling and non-Testardi behavior is expected at $T_{c}$.

To understand our results close to $T_{c}$ we determine the dynamic
spin response in the normal and SC phase. Analyzing the spin dynamics,
ballistic for $\omega\ll\Delta$ and overdamped for $\omega\gg\Delta$,
we find that the frequency integral in $\chi_{0,\mathrm{nem}}$ is
dominated by overdamped dynamics ($z=2$). Therefore, the key impact
of the SC phase is not on the dynamics, but on the static part of
the susceptibility. Magnetism in the pnictides has been shown to be
strongly affected by SC, to the extent that the AFM order parameter
is suppressed below $T_{\mathrm{c}}$ \cite{pratt,Christianson09}.
A Landau expansion for the competition of the AFM and SC order parameter
is governed by the coupling $\frac{\lambda_{m}}{2}\left\vert \Psi\left(x\right)\right\vert ^{2}\left(\mathbf{m}_{1}^{2}\left(x\right)+\mathbf{m}_{2}^{2}\left(x\right)\right)$
\cite{Sachdev04,Fernandes09}. Close to $T_{c}$ this coupling leads
to an increase in the inverse magnetic correlation length $r\rightarrow r+\lambda_{m}\left\langle \left\vert \Psi\left(x\right)\right\vert ^{2}\right\rangle $,
which suppresses the static part of the spin susceptibility $\tilde{\chi}_{p}$.
Thus, magnetic order and correlations are weakened in the SC state
and $\chi_{0,\mathrm{nem}}\left(0\right)=N\int_{p}\tilde{\chi}_{p}^{2}$
acquires a minimum at $T_{c}$. Given the overdamped spin dynamics
we obtain $\chi_{0,\mathrm{nem}}\left(0\right)\propto-\ln\left(r+\lambda_{m}\left\langle \left\vert \Psi\right\vert ^{2}\right\rangle \right)$
for $2D$ magnetic fluctuations. To show that this effect explains
the observed anomaly of $C_{\mathrm{s}}$ at $T_{\mathrm{c}}$, we
compare in Fig. 3 the RUS data for the optimally doped sample with
our theory. In general, the correction to $\xi^{-2}$ is proportional
to the composite operator $\left\langle \left\vert \Psi\right\vert ^{2}\right\rangle \propto\left(T_{c}-T\right)^{1-\alpha}$,
yet the SC transition is well described by a mean-field approach ($\alpha=0$).
On general grounds, one expects the AFM fluctuations to get weaker
along the overdoped region, which would lead to a less pronounced
softening of the lattice for $T>T_{c}$.

\begin{figure}

\begin{centering}
\includegraphics[width=0.8\columnwidth]{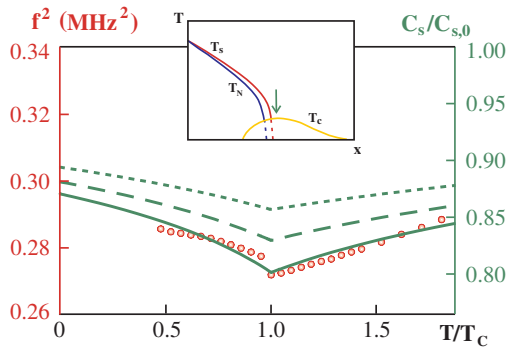} 
\par\end{centering}

\caption{Shear modulus $C_{s}$ in the tetragonal, SC phase. The inset shows
a schematic $\left(x,T\right)$ phase diagram for Ba(Fe$_{1-x}$Co$_{x}$)$_{2}$As$_{2}$,
with the green arrow denoting optimal doping. In the main panel, the
solid line refers to the optimally doped system, whereas the dashed
and dotted curves refer to systems that are respectively deeper in
the overdoped region. The red points refer to the RUS data of Fig.
2b.}

\end{figure}

We note that the reduction of $C_{s}$ close to the Tet-Ort transition
is a general feature, expected in any critical theory. 
However, the hardening of the lattice below $T_c$, where there is no structural instability,
is an unexpected observation. Yet, it finds a natural explanation in terms of
magnetically driven nematic fluctuations and competing AFM and SC order. 
Furthermore, the same model is able to explain the reduction of orthorhombic (i.e.
nematic) order in the SC state, as observed in Ref. \cite{Nandi09}. 
The existence of nematic fluctuations is crucial to our results: if one considered only
generic magnetic fluctuations, then the relevant magneto-elastic coupling would be
$\mathbf{m}^2 \varepsilon_{ii}$, instead of Eq. \ref{shearcoupling}. In this case, the magnetization would couple not to the shear distortion, 
but to the longitudinal modes, which cannot explain the strong correlation between the AFM and the Tet-Ort transitions.

In summary, the spin ordering in the pnictides results in emergent
nematic degrees of freedom that can be probed through the shear modulus.
Our results show that both magnetic and nematic/shear degrees of freedom are important low
energy excitations in a large temperature and doping regime of the
system. 

We thank S. Bud'ko, P. Canfield, A. Goldman, A. Kreyssig
and R. McQueeney for fruitful discussions. Work at the Ames Lab. and
ORNL was supported by the US DOE, Office of BES, under contracts No.
DE-AC02-07CH11358 and No. DE-AC05-00OR22725. L.H.V. and V.K. (NSF-DMR-0804719)
and P.C. (NSF-NIRT-ECS-0608842) thank the NSF for financial support.
The authors acknowledge the hospitality of the Aspen Center for Physics (J.S. and P.C.)
and of the Trinity College (S.B.).

\end{document}